\documentclass[%
 reprint,
 superscriptaddress,
 amsmath,amssymb,
 aps,
]{revtex4-2}

\usepackage{graphicx}
\usepackage{dcolumn}
\usepackage{bm}
\usepackage{lipsum}
\usepackage{siunitx}
\usepackage{hyperref}
\usepackage{placeins}
\usepackage{etoolbox}
\usepackage{xcolor}

\newcommand{\biscco}{$\text{Bi}_{2}\text{Sr}_{2}\text{CuCa}_{2}\text{O}_{8+\delta}$}
\DeclareSIUnit\bar{bar}

\makeatletter
\def\@email#1#2{%
 \endgroup
 \patchcmd{\titleblock@produce}
  {\frontmatter@RRAPformat}
  {\frontmatter@RRAPformat{\produce@RRAP{#1\href{mailto:#2}{#2}}}\frontmatter@RRAPformat}
  {}{}
}%

\begin{document}

\title{Encapsulating high-temperature superconducting twisted van der Waals heterostructures blocks detrimental effects of disorder}

\author{Yejin Lee}
 \affiliation{Leibnitz Institute for Solid State and Materials Science Dresden (IFW Dresden), 01069 Dresden, Germany}
 \affiliation{Institute of Applied Physics, Technische Universität Dresden, 01062 Dresden, Germany}
 \email{E-mail: y.lee@ifw-dresden.de} 
 
\author{Mickey Martini}
 \affiliation{Leibnitz Institute for Solid State and Materials Science Dresden (IFW Dresden), 01069 Dresden, Germany}
 \affiliation{Institute of Applied Physics, Technische Universität Dresden, 01062 Dresden, Germany}
 
  \author{Tommaso Confalone}
 \affiliation{Leibnitz Institute for Solid State and Materials Science Dresden (IFW Dresden), 01069 Dresden, Germany}
 \affiliation{Department of Physics, Sapienza University of Rome, 00185 Rome, Italy}
 
\author{Sanaz Shokri}
 \affiliation{Leibnitz Institute for Solid State and Materials Science Dresden (IFW Dresden), 01069 Dresden, Germany}
 \affiliation{Institute of Applied Physics, Technische Universität Dresden, 01062 Dresden, Germany}
 
  \author{Christian~N.~Saggau}
 \affiliation{Leibnitz Institute for Solid State and Materials Science Dresden (IFW Dresden), 01069 Dresden, Germany}

\author{Daniel Wolf}
\affiliation{Leibnitz Institute for Solid State and Materials Science Dresden (IFW Dresden), 01069 Dresden, Germany}
 
   \author{Genda Gu}
 \affiliation{Condensed Matter Physics and Materials Science Department, Brookhaven National Laboratory, Upton, NY 11973, USA}
 
   \author{Kenji Watanabe}
 \affiliation{Research Center for Functional Materials, National Institute for Materials Science, 1-1 Namiki, Tsukuba 305-0044, Japan}
 
    \author{Takashi Taniguchi}
\affiliation{International Center for Materials Nanoarchitectonics, National Institute for Materials Science,  1-1 Namiki, Tsukuba 305-0044, Japan}
    
 \author{Domenico Montemurro}
 \affiliation{Department of Physics, University of Naples Federico II, 80125 Naples, Italy}

  \author{Valerii~M.~Vinokur}
 \affiliation{Terra Quantum AG, Kornhausstrasse 25,
CH-9000 St.\,Gallen, Switzerland}

 \author{Kornelius Nielsch}
 \affiliation{Leibnitz Institute for Solid State and Materials Science Dresden (IFW Dresden), 01069 Dresden, Germany}
 \affiliation{Institute of Applied Physics, Technische Universität Dresden, 01062 Dresden, Germany}
 \affiliation{Institute of Materials Science, Technische Universität Dresden, 01062 Dresden, Germany}
 
 \author{Nicola Poccia }
 \affiliation{Leibnitz Institute for Solid State and Materials Science Dresden (IFW Dresden), 01069 Dresden, Germany}


\begin{abstract}

High-temperature cuprate superconductors based van der Waals (vdW) heterostructures hold high technological promise. One of the obstacles hindering the progress is the detrimental effect of disorder on the properties of the vdW devices-based Josephson junctions (JJ). Here we report the new method of fabricating twisted vdW heterostructures made of \biscco\,, crucially improving the JJ characteristics and pushing them up to those of the intrinsic JJs in bulk samples. The method combines cryogenic stacking using a solvent-free stencil mask technique and covering the interface by insulating hexagonal boron nitride crystals.
Despite the high-vacuum condition down to $\SI{e-6} {\milli\bar}$ in the evaporation chamber, the interface appears to be protected from water molecules during the in-situ metal deposition only when fully encapsulated. Comparing the current-voltage curves of encapsulated and unencapsulated interfaces, we reveal that the encapsulated interfaces’ characteristics are crucially improved so that the corresponding JJs demonstrate high critical currents and sharpness of the superconducting transition comparable to those of the intrinsic JJs. Finally, we show that the encapsulated heterostructures are more stable over time.

\end{abstract}

\maketitle

\onecolumngrid
\twocolumngrid

\section{Introduction}

Layered quasi-two dimensional (2D) materials comprising the stack of monolayers held together by van der Waals (vdW) forces can be cleaved via a simple scotch tape exfoliation down to constitutive monolayers\,\cite{novoselov2005two}. High temperature superconductors (HTSC) provide a wide variety of such layered correlated systems. Remarkably, even the atomically thin\,\biscco (BSCCO) layers, i.e., the layers containing a single or a few elementary cells, have been found to possess the superconducting transition temperature close to that of the bulk samples\,\cite{yu2019high, zhao2019sign} and showed the superconductor-insulator transition driven by the evolution of the density of states\,\cite{liao2018superconductor}. Because of these properties, HTSCs can serve as starting building blocks for the vdW heterostructures. However, isolating the cuprate single layers that hold superconductivity remains a challenging task, especially if one wishes to realize thin and crystalline-ordered interfaces. The point is that the atomically thin BSCCO flakes turn highly insulating if contaminated with oxygen under the ambient atmosphere\,\cite{novoselov2005two, sandilands2014origin}. Raman measurements\,\cite{sandilands2014origin, sandilands2010stability} reported high chemical activity of oxygen in thin BSCCO flakes. More detailed studies\,\cite{huang2022unveiling} revealed that water molecules can also quickly deteriorate the surface of BSCCO flakes. In addition, oxygen dopants in cuprates are mobile above 200\,K\,\cite{poccia2011evolution, fratini2010scale}, destroying high-quality superconductivity requiring ordered distribution of oxygen defects\,\cite{poccia2012optimum, campi2015inhomogeneity}. In comparison with the bulk crystalline order, the robustness of the spatially-correlated superlattice orders in BSCCO down to a few unit cells is remarkable\,\cite{poccia2020spatially}. Thus, cryogenic temperatures and a well-controlled environment are necessary to prevent detrimental disorder effects in the spatially correlated super-lattices and to freeze oxygen defects in their functional original positions for realizing the high quality cuprates-based vdW heterostructures.

The possibility of making twisted HTSC-based heterostructures has attracted substantial interest because of the $d$-wave pairing symmetry\,\cite{klemm2000order}. Previously twisted BSCCO junctions using the bulk crystals or flakes were realized through an annealing process at high temperature in oxygen atmosphere\,\cite{zhu2021presence, li1999bi}, which can reconstruct the interfacial structure\,\cite{zhu1998structural}. The obtained structures did not demonstrate any angular dependence of the Josephson current\,\cite{zhu2021presence,li1999bi}. Nevertheless, the non-monotonic angular behavior of the critical current has been reported on the cross-whisker HTSCs junctions, with the critical current being much reduced as compared to the critical current of the bulk intrinsic junctions\,\cite{latyshev2004c}. On the other hand, the angular dependence of the critical current in cuprate in-plane grain boundary junctions is indeed well described by the $d$-wave pairing symmetry because of the large in-plane coherence length as compared to that in the out-of-plane junctions\,\cite{mannhart2002experiments,hilgenkamp2002grain}, which reduces the detrimental disorder effects. However, the transmission electron microscopy has shown that the grain-boundary Josephson junctions (GBJJs) are generally composed of facets in the range of 10–100\,nm and demonstrate a strong dependence of their properties upon the particular HTSC, the substrate, the conditions of the film deposition, and of the presence of defects\,\cite{PhysRevB.95.184505}. Facets could take place in all three dimensions as a consequence of the adopted fabrication techniques for bicrystals, biepitaxial growth, or step edges\,\cite{hilgenkamp1996implications}. The GBJJs have been well studied, and their electronic properties were found to be controlled by the misorientation between two grains\,\cite{dimos1988orientation}. Because of that, the facets create additional complexity and difficulties in controlling the JJs' properties.

\begin{figure*}[t!]
    \center
    \includegraphics[width=0.95\textwidth]{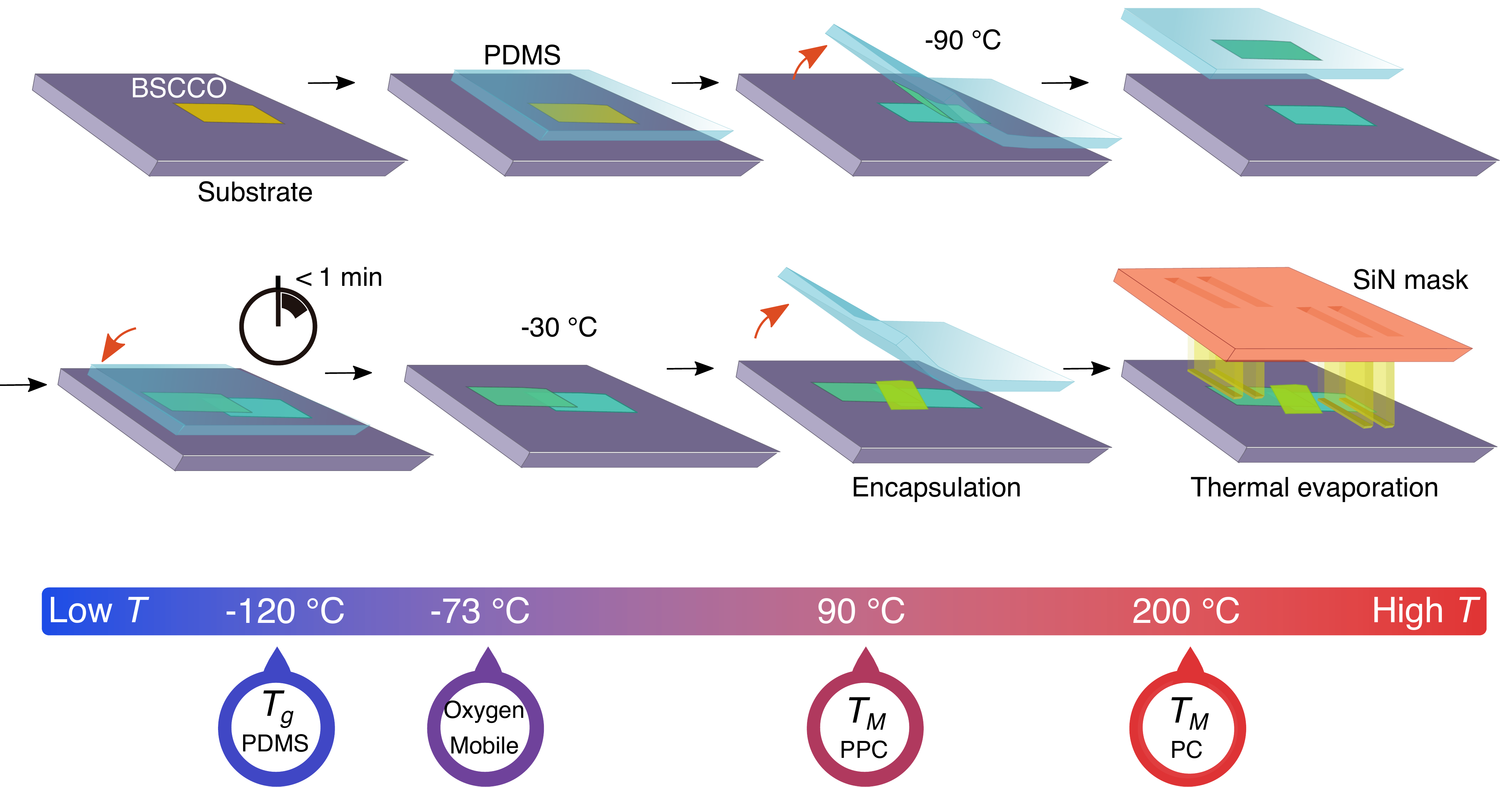}
    \caption{Cryogenic fabrication process for BSCCO junctions. An exfoliated BSCCO flake is identified by its optical contrast under a microscope and a PDMS stamp is approached towards the flake on the substrate. The cold PDMS stamp is dragged, cleaving the crystal at $-\SI{90}{\celsius}$ and then quickly aligned along the flake. Stacked flakes are slowly warmed to $-\SI{30}{\celsius}$, and the PDMS is pulled off. The junction interface is covered by landing an hBN flake using a PDMS at room temperature. Electrical contacts on the crystals are deposited by thermal evaporation through a SiN stencil mask. The color bar below indicates representative temperatures for oxygen in cuprates and different types of polymer(PDMS, PPC, and PC).}
    \label{fig:FIG1}
\end{figure*}

An angular dependence of the critical current of an out-of-plane Josephson junction resulting in its change over two orders of magnitude has been demonstrated in thin BSCCO twisted heterostructures prepared by the cryogenic stacking technique while preserving the coherence of the crystalline and oxygen order at the interface\,\cite{zhao2021emergent}. An additional important evidence of the significant reduction of the Josephson critical current when changing the twist angle has been reported in\,\cite{lee2021twisted}. It was found that the corresponding critical currents are, on average, lower than the critical current of an intrinsic Josephson junction for the BSCCO\,\cite{zhao2021emergent}, given that the fabrication did not occur under cryogenic conditions. The general improvement of the control over the BSCCO properties in the low-dimensional limit stimulated the theoretical activities. The emergence of the topological states in the twisted vdW heterostructures of HTSC BSCCO layers with the $d$-wave superconducting order parameter was suggested\,\cite{can2021probing, tummuru2022josephson, tummuru2022twisted, mercado2022high, cadorim2022vortical,can2021high}. The twist angle close to $\SI{45}{\degree}$ was found to result in a time-reversal symmetry (TRS) broken chiral superconducting $d_{x^2-y^2}\pm id_{xy}$ phase, which was also reported at the intermediate twist angles and was attributed to the unconventional sign structure of the $d$-wave order parameter\,\cite{tummuru2022josephson, volkov2021josephson}. Strong support for this theoretical proposal came from the experimental detection of some new interfacial superconductivity\,\cite{zhao2021emergent}, manifesting as a dominant second harmonic of the Josephson current close to $\SI{45}{\degree}$ angle. However, the TRS breaking in the high temperature superconducting phase can be suppressed by strong disorder at the interface\,\cite{volkov2021josephson}, hence careful studies of detrimental disorder effects on the interfaces and novel methods that rely on cheaper and/or innovative process of fabrication are required.
\\
\newpage

\section{Fabrication}

\begin{figure*}[t!]
 \centering
 \includegraphics[width=.78\textwidth]{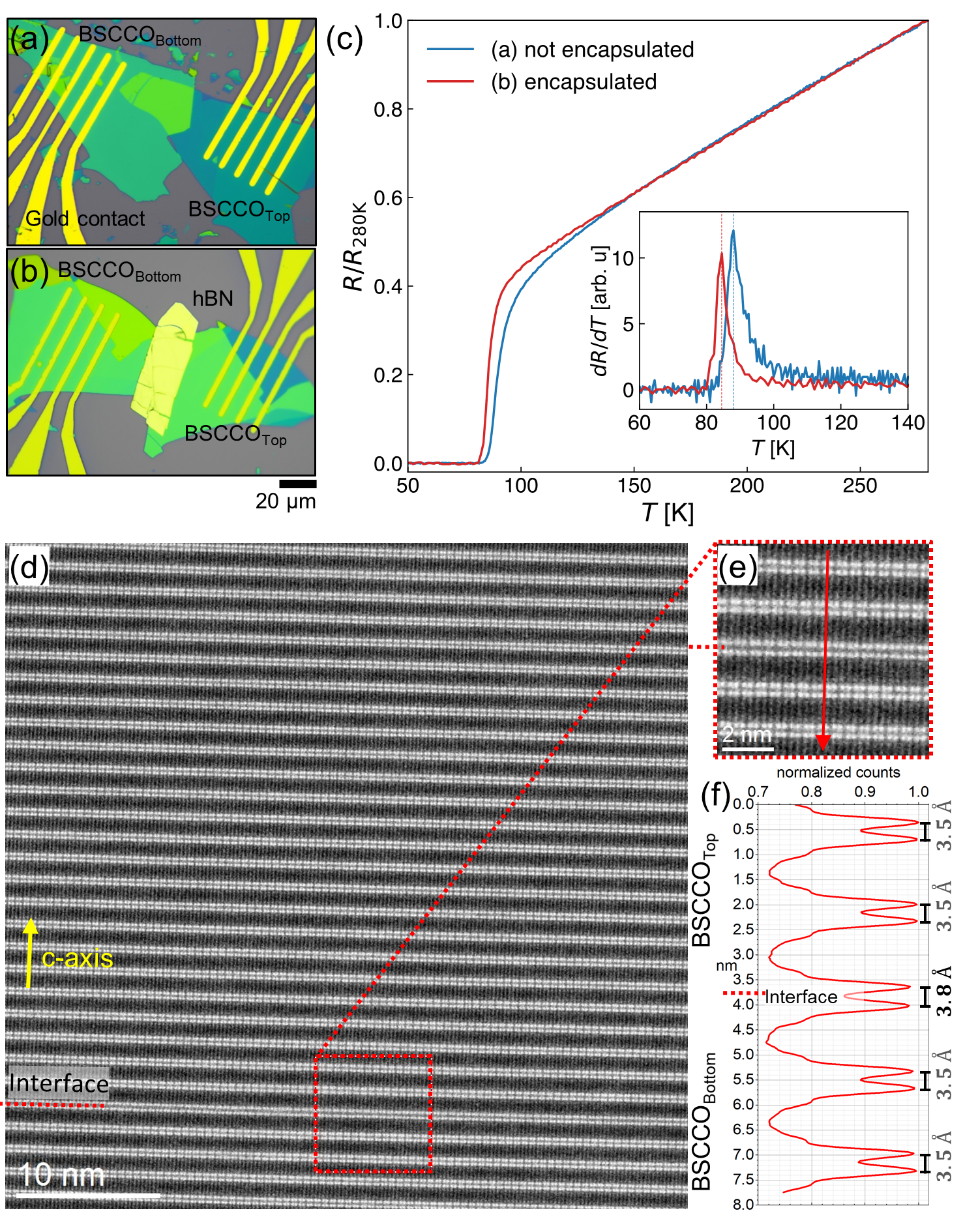}
 \caption{(a) Optical micrograph of BSCCO junctions with the uncovered interface and (b) hBN-encapsulated interface. (c) Normalized resistance $R/R_{280 \mathrm{K}}$ as a function of temperature $T$ of corresponding BSCCO junctions in (a) and (b), respectively. Inset: The differential resistance $dR/dT$ as a function of temperature. The dotted lines show superconducting critical temperature $T_{c}$ for junctions with encapsulation (red) and without encapsulation (blue). (d) Cross-sectional high-angle annular dark-field scanning transmission electron microscopy (HAADF-STEM) image of a BSCCO junction recorded in [100] zone axis orientation. The Bi atoms are only atomically resolved. (e) Zoom-in at the junction indicated by the red dashed box emphasizing the high quality of the junction between the flakes. (f) The line profile across the junction at the position indicated by a red arrow in (e) but averaged over the whole width of the image (d) reveals a significantly larger distance between the Bi atoms at the junction (3.8 Å) compared to the Bi atoms (3.5 Å) at each flake along the c-axis.}
 \label{fig:FIG2}
\end{figure*} 

\begin{figure*}[t!]
 \centering
 \includegraphics[width=.86\textwidth]{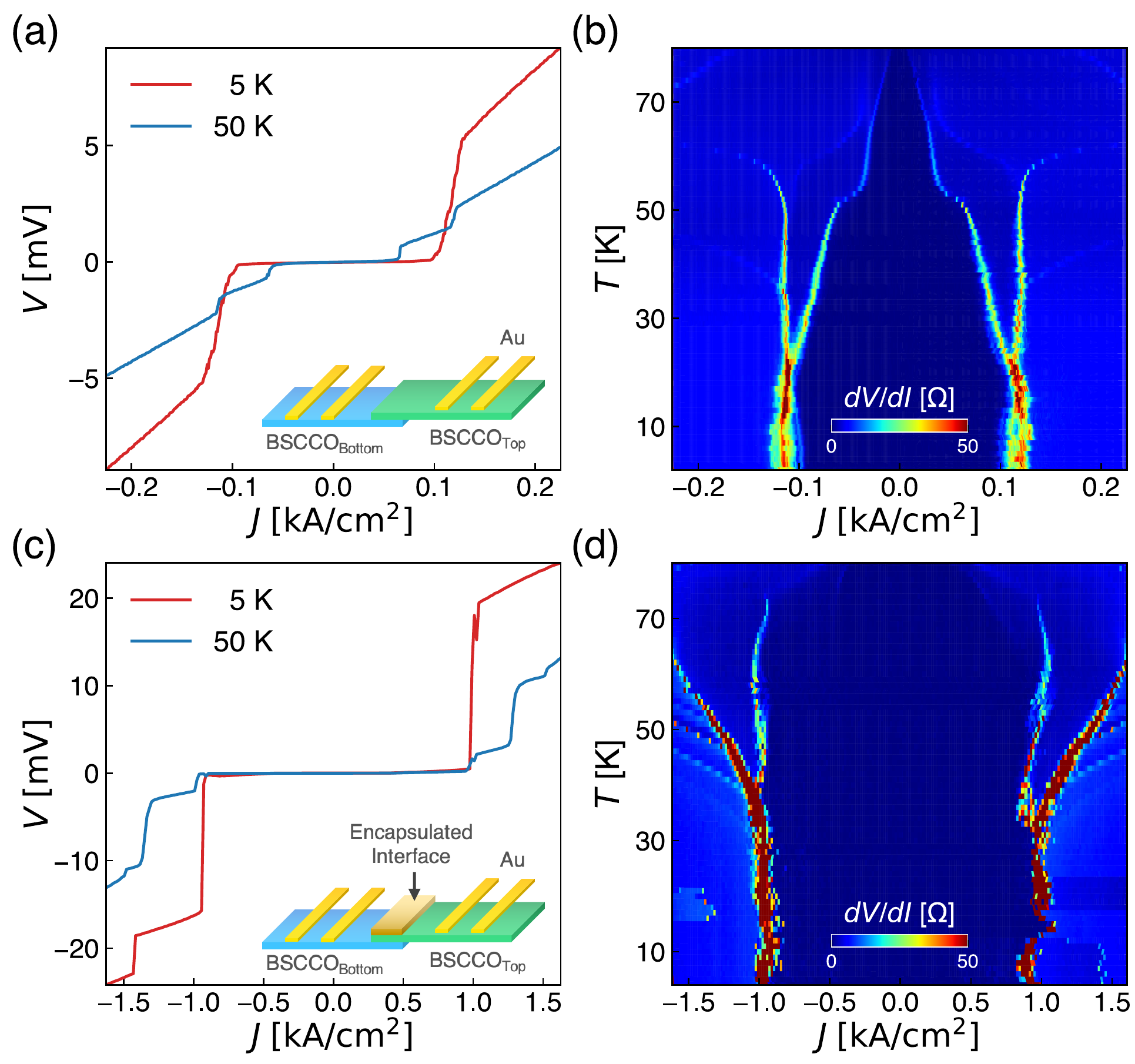}
 \caption{(a) Representative current density-voltage $J$-$V$ characteristics at $\SI{5}{\kelvin}$ and $\SI{50}{\kelvin}$ Inset: Schematic illustration of corresponding BSCCO junction. (b) Differential resistance as a function of current density and temperature of the BSCCO junction without encapsulation on the interface. (c) Analogous current density-voltage characteristics at $\SI{5}{\kelvin}$ and $\SI{50}{\kelvin}$. Inset: Schematic illustration of encapsulated BSCCO junction. (d) Differential resistance versus current density and temperature of the BSCCO junction with the encapsulated interface.}
 \label{fig:FIG3}
\end{figure*} 

We fabricate six $\SI{0}{\degree}$ Josephson junctions and one $\SI{43.2}{\degree}$-twisted Josephson junction based on the optimally doped BSCCO flakes using a cryogenic dry transfer technique in a pure argon atmosphere. This technique consists of cleaving a sequential pair of fresh surfaces of the BSCCO from the pre-exfoliated single crystal and stacking the two resulting flakes on top of each other. This procedure of the junction fabrication is sketched in Fig.\,\hyperref[fig:FIG1]{1} and can be described as follows. First, the BSCCO crystals are mechanically exfoliated using a scotch tape on the $\text{SiO}_{2}/\text{Si}$ substrates, previously treated with oxygen plasma and baked overnight to get rid of water molecules. Next, utilizing the optical contrast, we identify the BSCCO flake with a thickness in the range between 80 and $\SI{100}{\nano\meter}$ and cool the sample stack down to $-\SI{90}{\celsius}$, to preserve the crystalline structure and the superconducting state of the interface while building the junction. After that, we cover the BSCCO flake with an edge of a polydimethylsiloxane (PDMS) stamp placed on a glass slide mounted on a micromanipulator and let the assembly thermalize. At the temperature, being a little bit above the glass transition of our PDMS ($T_g = -\SI{120}{\celsius}$), the stamp becomes very adhesive. By quickly detaching the stamp from the substrate, we cleave the crystal along a flat plane between the BiO planes, obtaining two thinner flakes. The flake standing on the PDMS stamp is aligned and placed back onto the bottom flake on the substrate within 40 seconds. The time between cleaving and stacking makes a huge impact on the junction quality. We find out that both flakes should be thicker than $\SI{30}{\nano\meter}$, otherwise they are not rigid enough to create a flat interface without modulating the surface. Finally, the stack is slowly heated up to $-\SI{30}{\celsius}$ and the top flake is released from the bottom one as the stamp is no longer adhesive. In Figure\,\hyperref[fig:FIG1]{1}, the bottom color bar illustrates representative temperatures (eg. melting temperature $T_{M}$) of the commonly used polymers for exfoliation, such as polycarbonate(PC) and polypropylene carbonate(PPC)\cite{wang2013one}. The control over the adhesion of the PDMS stamp allows us to fabricate the junctions by taking advantage of a solvent-free and dry transfer at low temperatures. In this way, the encapsulated sample survives after being exposed to an ambient condition for at least three hours.

For three out of the seven junctions (two at \SI{0}{\degree}, and one twisted at $\SI{43.2}{\degree}$), we opt to additionally protect the interface above the BSCCO heterostructures, especially from water molecules by placing an encapsulating hexagonal boron nitride (hBN) flake on top of the stacked flakes immediately afterward. The bottom surface of the lower flake in the heterostructure is attached closely to the substrate and, therefore, is not exposed to the ambient atmosphere, which is critical for the degradation of BSCCO. Electrical contacts are then deposited in two steps using a chemical-free stencil mask technique\,\cite{zhao2019sign} in an evaporation chamber directly connected to the glovebox cluster. First, gold electrodes are evaporated right on the junctions while the temperature of the liquid nitrogen-cooled stage is kept at $-\SI{50}{\celsius}$. The base pressure of the deposition chamber is $\SI{e-6}{\milli\bar}$. Next, electrical contact pads are deposited via stencil masks by evaporating Au/Cr while the temperature of the sample stage is $-\SI{30}{\celsius}$. These double evaporations aim to avoid deposition of the Cr directly onto the BSCCO, which would result in an insulating behavior of the underlying region, as Cr could capture oxygen atoms from the flakes\,\cite{ghosh2020demand}. We found that the contact resistance obtained from the deposition of Cr right onto the sample by performing a single evaporation (Au/Cr) was of the order of $\SI{1}{\kilo\ohm}$, which is ten times higher than the resistance of the contact obtained by the two-steps evaporation protocol.
Figures \ref{fig:FIG2}(a) and (b) show the closeup optical images of two representative devices, one without encapsulation on top of the junction and the other with encapsulation, respectively. Figure\,\hyperref[fig:FIG2]{2(c)} presents the temperature dependence of the electrical resistance $R$ through the junctions (normalized by the resistance at $\SI{280}{\kelvin}$) of the two corresponding devices. The $R(T)$ of the interface is measured across the junction with the four-terminal method. In addition, the resistance as a function of temperature for top and bottom flakes is reported in the supplementary information in Fig. S2. The $T$ dependence of $R$ is linear in the normal state, consistent with the nearly optimal doped \biscco\,\cite{qiu2011high} and exhibits a superconducting transition, at a temperature slightly lower than that of the bulk samples\,\cite{wen2008large}. The superconducting critical temperature $T_{c}$ in the junction with the encapsulated (uncovered) interface is $\SI{84.5}{\kelvin}$ ($\SI{88}{\kelvin}$), defined as the peak value in the derivative $dR/dT$ (Fig.\,\hyperref[fig:FIG2]{2(c)}). The $T_{c}$ is close to that of a bulk crystal, indicating the high uniformity of oxygen dopants even at the junction in both of the devices, and its values are within the distribution of $T_{c}$ in twisted BSCCO junctions\cite{zhao2021emergent}. Yet, the device with the encapsulated junction displays a visibly sharper transition from the resistive to the metallic state, indicating the improved interplanar coupling\,\cite{kao1993systematics}. The critical temperatures $T_c$ of other JJs are given in Tab.\,\ref{tab:table} along with the other characteristic properties. The resistances as functions of temperature for all ${\SI{0}{\degree}}$ junctions are displayed in Fig.\,S1.

Cross-sectional high-annular dark-field scanning transmission electron microscopy (HAADF-STEM) is performed to image the $\SI{0}{\degree}$ BSCCO junction at atomic resolution. We prepare a lamella cut out of the BSCCO junction by focused ion beam milling. The TEM measurement is carried out immediately after the specimen preparation to prevent deterioration from air exposure. Moreover, the sample is treated with Ar-O plasma for $\SI{30}{\second}$ to suppress contamination while imaging locally with an electron beam. Figure\,\hyperref[fig:FIG2]{2(d)} illustrates the HAADF-STEM image of the junction's cross-section between the top and bottom flakes, recorded in [100] zone axis orientation. Bright spots are identified as atoms that scatter electrons, and the brightest ones are Bi atoms, which are only fully resolved in these imaging conditions. Figure\,\hyperref[fig:FIG2]{2(e)} shows zoom-in at the junction indicated by the red dashed box in Fig.\,\hyperref[fig:FIG2]{2(d)}, emphasizing the high quality of the junction between the flakes. The monolayers of BSCCO are placed in series along the c-axis with the lattice constant of around $\SI{1.5}{\nano\meter}$ for both flakes and the crystalline order in the junction is well preserved. The line profile across the junction sketched by a red arrow is displayed in Fig.\,\hyperref[fig:FIG2]{2(f)}, presenting the larger distance between the Bi atoms at the junction ($\SI{3.8}{\angstrom}$) than that of Bi atoms in each crystal ($\SI{3.5}{\angstrom}$) along the c-axis. The c-axis position of the interface is in an agreement with the thickness of the two flakes estimated by its optical contrast. The STEM image of the full stack for the junction is shown in Fig S4.

\section{Results and Discussion}

\begin{table}[h]
    \centering
    \caption{Summary of the properties of the twisted Josephson junctions.}

    \begin{tabular}{cccccc}
         Sample &$\theta [\SI{}{\degree}]$  &Encapsulation &Area$  [\SI{}{\micro\meter^2}]$ &$T_c [\SI{}{\kelvin}]$ &$J_c @ \SI{10}{\kelvin}  [\SI{}{\kilo\ampere\per\centi\meter^2}]$\\

         \hline
         S1 &0.0 &no &670 &88.5 &0.10 \\ 
         S2 &0.0 &no &640 &84.5 &0.21\\ 
         S3 &0.0 &no &240 &89.0 &0.38 \\ 
         S4 &0.0 &no &386 &89.0 &0.65 \\
         S5 &0.0 &yes &120 &84.5 &1.00\\ 
         S6 &0.0 &yes &300 &88.7 &0.94 \\ 
         S7 &43.2 &yes &160 &91.5 &0.04
    \end{tabular}
    \label{tab:table}
\end{table}

\begin{figure*}[t!]
\centering
\includegraphics[width=0.87\textwidth]{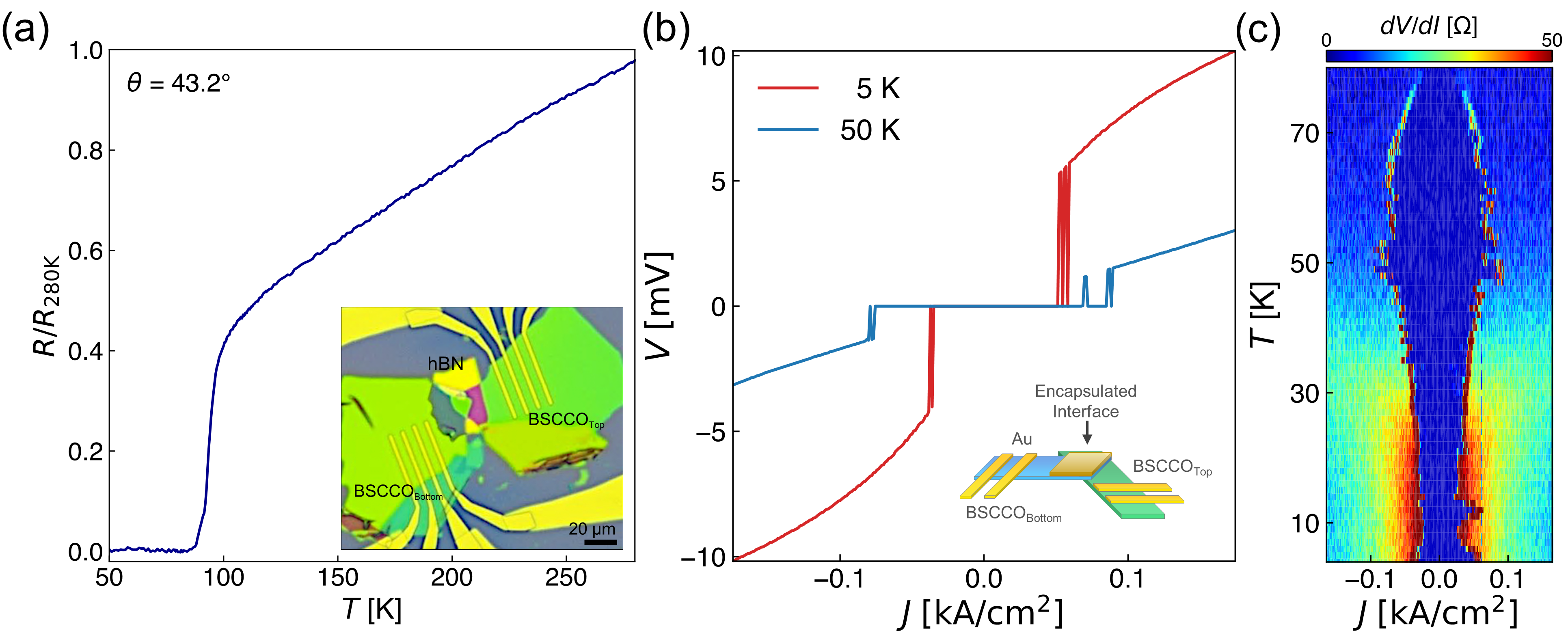}
\caption{(a) Temperature dependence of the resistance through \SI{43.2}{\degree}-twisted Josephson junction with hBN-encapsulation. Inset: Optical image of the corresponding device. (b) Representative current density-voltage $J$-$V$ characteristics at $\SI{5}{\kelvin}$ and $\SI{50}{\kelvin}$. Inset: Schematic illustration of corresponding BSCCO junction. (c) Differential resistance as a function of current density and temperature for the \SI{43.2}{\degree}-twisted Josephson junction.}
\label{fig:FIG4}
\end{figure*}

The current-voltage, $I$-$V$, characteristics and the dynamical resistance, $dV/dI$, across the JJs are measured in a four-terminal configuration using a lock-in amplifier technique. We sweep the dc current from negative to positive values in the range of $\SI{}{\milli\ampere}$ while superimposing the small alternating current with an $\SI{1}{\micro\ampere}$-amplitude and a frequency of $\SI{17}{\hertz}$. We measure simultaneously the dc and ac voltages to get the $I$-$V$ and $dV/dI$ characteristics, respectively. 
To compare the transport data of the two representative JJs (one uncovered and the other encapsulated), we normalize the bias current $I$ with respect to the junction area $A$, obtaining the current density $J = I/A$. Figure\,\hyperref[fig:FIG3]{3(a)} displays the representative $J$-$V$ characteristics of the non-encapsulated junction shown in Fig.~\hyperref[fig:FIG2]{2(a)} at $\SI{5}{\kelvin}$ and $\SI{50}{\kelvin}$. The schematic of the device is sketched in the inset of Fig.\,\hyperref[fig:FIG3]{3(a)}. As we sweep the bias current from a large negative value, the junction voltage first retraps to the superconducting state, $V=0$, and then jumps to the resistive state at the critical current density $J_c$. In this JJ, $J_c = \SI{0.10}{\kilo\ampere\per\centi\meter^2}$ at $\SI{5}{\kelvin}$ and decreases to $J_c = \SI{0.06}{\kilo\ampere\per\centi\meter^2}$ at $\SI{50}{\kelvin}$. 
Figure\,\hyperref[fig:FIG3]{3(b)} shows the differential resistance $dV/dI$ as a function of the current density and temperature. The critical current density $J_c$, at which the inner peaks of $dV/dI$ appear, decreases with temperature. Furthermore, multiple peaks are visible at higher bias current (for both bias polarities) above $\SI{25}{\kelvin}$ approximately. These features possibly arise from the subgap structure of the quasiparticle density of states in the intrinsic BSCCO Josephson junctions\,\cite{schlenga1996subgap,itoh1997current}. The $dV/dI$ is symmetric with respect to $J=0$.

The $J$-$V$ characteristics of the encapsulated JJ depicted in Fig.\,\hyperref[fig:FIG2]{2(b)} measured at $\SI{5}{\kelvin}$ and $\SI{50}{\kelvin}$ are presented in Fig.\,\hyperref[fig:FIG3]{3(c)}. The sketch of corresponding device is displayed in the inset of Fig.\,\hyperref[fig:FIG3]{3(d)}. The transition from the superconducting to resistive state occurs at the current density almost ten times higher than that of the former junction ($J_{c}=\SI{1.19}{\kilo\ampere\per\centi\meter^2}$ at $\SI{5}{\kelvin}$). This value is comparable to the $J_c$ of the intrinsic BSCCO Josephson junctions, which range from $\SI{0.17}{}$ to $\SI{1.70}{\kilo\ampere\per\centi\meter^2}$ at $\SI{10}{\kelvin}$, depending on the number of junctions along the $c$-axis\,\cite{irie2000critical}. This JJ exhibits a sharp transition where the system rapidly jumps at $J_{c}$ from zero voltage to a finite voltage of $\SI{20}{\milli\volt}$ at $\SI{5}{\kelvin}$, while it ends up to a voltage of $\SI{2}{\milli\volt}$ at $\SI{50}{\kelvin}$. This suggests that the superconducting gap decreases by order of magnitude when the temperature raises from $\SI{5}{\kelvin}$ to $\SI{50}{\kelvin}$, as the switching voltage corresponds to the superconducting gap values in ideal tunnel junctions \cite{barone1982physics}. Furthermore, the $J$-$V$ curve at $\SI{50}{\kelvin}$ displays additional voltage jumps when the current density exceeds the $J_{c}$. This suggests the contribution of each intrinsic junction in crystals in the framework of the resistively and capacitively shunted junction (RCSJ) model using two resistance in series.\cite{chana2000alternating}. 

To gain further insight, we plot in Fig.\,\hyperref[fig:FIG3]{3(d)} the $dV/dI$ of this encapsulated JJ versus current density and temperature. The first peaks of $dV/dI$ occurring at $J_c$ exhibit a non-monotonic behavior as a function of temperature. In this case, $J_{c}$ is not mirrored along $J=0$ below $\SI{40}{\kelvin}$, and small hysteretic behavior is presented. Namely, the current density at which the junction retraps to the zero resistance state from the negative bias is slightly lower than the current density at which the system jumps to the resistive state at the positive bias. Above $\SI{40}{\kelvin}$, $dV/dI$ is symmetric and several peaks can be identified above $J_c$.
These features appear at different current densities depending on the temperature and can be associated either with the subgap structure as in the previous case\,\cite{schlenga1996subgap,itoh1997current}, the formation of the discrete vortex train\,\cite{zybtsev2005switching}, or with the multiparticle tunneling mechanisms\,\cite{schrieffer1963two}, such as phonon-assisted tunneling process\,\cite{schlenga1998tunneling,wolf2011principles}. These structures in the subgap regime are well explained by the resonant coupling mechanism between the infrared active optical $c$-axis phonons and oscillating Josephson currents\,\cite{schlenga1998tunneling}. 
When the bias is much higher than the critical current, no structures are presented, and the $J$-$V$ characteristic is linear.

The dynamical resistance over current density and temperature for all the six $\SI{0}{\degree}$ samples are shown in Fig. S3. Comparing the encapsulated junction (S6) with the similar area as the ones without encapsulation (S3 and S4), S6 with the area of $\SI{300}{\micro\square\meter}$ has the $J_c$ of $\SI{0.94}{\kilo\ampere\per\centi\meter^2}$ at $\SI{10}{\kelvin}$, which is higher than the others. In addition, its value of $J_c$ is reduced compared to that of the encapsulated junction with the smaller area (S5). This suggests the hBN protects the interface regardless of the area.

The experimental method combining cryogenic dry transfer technique and hBN encapsulation is also employed to build the twisted JJ with the twist angle $\theta = \SI{43.2}{\degree}$, aiming at the investigation of the anisotropy in the superconducting order parameter of BSCCO. The optical image of this device is shown in Fig.\,\hyperref[fig:FIG4]{4(a)} along with the temperature dependence of the normalized electrical resistance measured across the junction in a four-point geometry. Although the resistance curve is indistinguishable from that of the $\SI{0}{\degree}$-junctions [Fig.\,\hyperref[fig:FIG2]{2(c)}], the $J$-$V$ characteristic is distinctive. The two $J$-$V$ curves in Fig.\,\hyperref[fig:FIG4]{4(b)} at $\SI{5}{\kelvin}$ and $\SI{50}{\kelvin}$ display a significant steep jump at the critical current density below $\SI{0.1}{\kilo\ampere\per\centi\meter^2}$. As visible also in the $dV/dI$ colormap [Fig.~\hyperref[fig:FIG4]{4(c)}], the $J_c$ is reduced from the value it has in the intrinsic junction by almost two orders of magnitude. Here, the drop of $J_c$ is not due to the detrimental effects of disorder since the interface is protected by the encapsulation layer. This conclusion is clearly supported by the sharp transition shown by the $J$-$V$ curves [Fig.~\hyperref[fig:FIG4]{4(b)}]. The reduction of $J_c$ is attributed instead to the $d$-wave pairing symmetry of the BSCCO. Near $\SI{45}{\degree}$, the nodal SOPs of the two twisted crystals strongly mismatch, and the direct tunneling of Cooper pairs is strongly suppressed, resulting in a lower value of $J_c$\,\cite{tummuru2022josephson, volkov2021josephson}.


\section{Conclusion}
In summary, we have investigated the detrimental effects of disorder on the electronic transport in JJs based on stacked BSCCO heterostructures and demonstrated the improvement of the corresponding $I$-$V$ characteristics and the $dV/dI$ due to encapsulated interface with an hBN-crystal, fabricated using cryogenic, solvent-free transfer techniques in Ar atmosphere. Encapsulation of the junction interface is, under our experimental conditions, of vital importance for the fabrication of JJs with an electronic quality comparable to that in the intrinsic Josephson junctions in the single-crystal BSCCO. Our main finding is the enhancement of $J_c$ in the encapsulated JJ, which becomes as high as $J_c$ in the intrinsic Josephson junction, achieving the values $J_c \sim \SI{1.2}{\kilo\ampere\per\centi\meter^2}$ at $\SI{10}{\kelvin}$, whereas $J_c$ in the not-encapsulated JJs is about one order of magnitude smaller. The sharper superconducting transition demonstrated by the $R(T)$ dependencies and the presence of structures in the $J$-$V$ characteristics observed in the encapsulated device do coincide with the main features of the coherent JJs free of detrimental disorder making the vdW heterostructures with the encapsulated interface an excellent platform for applications. 

\medskip
\noindent\textbf{Supporting Information.} \par 
\noindent Supporting Information is available from the Wiley Online Library.

\medskip
\noindent \textbf{Acknowledgements.} \par
\noindent The experiments were partially supported by the Deutsche Forschungsgemeinschaft (DFG 452128813, DFG 512734967, DFG 492704387, DFG 460444718). The work of V.M.V. was supported by Terra Quantum AG. The authors are grateful to Heiko Reith and Nicolas Perez Rodriguez for providing access to cleanroom and cryogenic facilities respectively. D.W. acknowledges funding from DFG SFB 1415, Project ID No. 417590517. K.W. and T.T. acknowledge support from the JSPS KAKENHI (Grant Numbers 19H05790, 20H00354, and 21H05233). The work at BNL was supported by the US Department of Energy, oﬃce of Basic Energy Sciences, contract no. DOE-sc0012704. The authors are also  grateful to Shu Yang Frank Zhao, Philip Kim, Francesco Tafuri, Giampiero Pepe; Naurang Lal Saini, and Valentina Brosco for illuminating and fruitful discussions. \\

\noindent\textbf{Author contributions.} \par
\noindent N.P. conceived and designed the experiment; Y.L., M.M. performed the experiments and analyzed the data with the contribution of T.C., S.S., and C.N.S. The cuprate crystals have been provided by G.G. The hexagonal boron nitride crystals have been provided by K.W. and T.T. The TEM measurement is carried out by D.W. The fabrication procedure and the results have been discussed by N.P., Y.L., M.M., and  V.M.V. The manuscript has been written by N.P., Y.L., M.M., D.M., V.M.V., and K.N. All authors discussed the manuscript.\\

\noindent\textbf{Conflict of Interest.} \par
\noindent All authors declare no conflict of interest.

\bibliographystyle{}

\end{document}